\documentclass[apj]{emulateapj}
\usepackage{apjfonts} \usepackage{amsmath} 
\newcommand{\me}{$M_{\oplus}$} 
\newcommand{\re}{$R_{\oplus}$}

\newcommand{\teq}{$T_{\rm eq}$}
\newcommand{\cp}{\citep} 
\newcommand{\ct}{\citet}

\shorttitle{Photospheric Radii of Exoplanets}
\shortauthors{Fortney et al.}

\begin{document}

\title{Exploring A Photospheric Radius Correction to Model Secondary Eclipse Spectra for Transiting Exoplanets}

\author{Jonathan J. Fortney\altaffilmark{1}, Roxana E. Lupu\altaffilmark{2}$^,$\altaffilmark{3}, Caroline V. Morley\altaffilmark{4}, Richard S. Freedman\altaffilmark{3}$^,$\altaffilmark{5}, Callie Hood\altaffilmark{1}}

\altaffiltext{1}{Department of Astronomy \& Astrophysics, University of California, Santa Cruz, USA 95064 jfortney@ucsc.edu}
\altaffiltext{2}{BAER Institute, Moffett Field, CA}
\altaffiltext{3}{Space Science and Astrobiology Division, NASA Ames Research Center, Moffett Field, CA, USA}
\altaffiltext{4}{Department of Astronomy, University of Texas, Austin}
\altaffiltext{5}{SETI Institute, Mountain View, CA, USA}

\begin{abstract} We highlight a physical effect that is often not considered that impacts the calculation of model spectra of planets at secondary eclipse, affecting both emission and reflection spectra.  The radius of the emitting surface of the planet is not merely one value measured from a transit light curve, but is itself a function of wavelength, yet it is not directly measurable.  At high precision, a similar effect is well-known in transit ``transmission spectroscopy'' but this related effect also impacts emission and reflection.  As is well-appreciated, the photospheric radius can vary across $\sim$4-8 atmospheric scale heights, depending on atmospheric opacity and spectral resolution.  This effect leads to a decreased weighting in model calculations at wavelengths where atmospheric opacity is low, and one sees more deeply into the atmosphere, to a smaller radius.  The overall effect serves to mute emission spectra features for atmospheres with no thermal inversion but to enhance features for atmospheres with a thermal inversion.  While this effect can be ignored for current \emph{Hubble} observations, it can lead to wavelength-dependent 10-20\% changes in planet-to-star flux ratios in the infrared at $R\sim~200-1000$ (readily achievable for JWST) for low-gravity hot Jupiters, although values of 5\% are more typical for the population.  The effect is mostly controlled by the ratio of the atmospheric scale height to the planet radius, and can be important at any planetary temperature.  Of known planets, the effect is largest for the cool ``super-puffs" at very low surface gravity, where it can alter calculated flux ratios by over 100\%.  We discuss complexities of including this photospheric radius effect in 1D and 3D atmosphere models.

\end{abstract}

\maketitle

\section{Introduction}\label{introduction}
When an exoplanet is occulted by its parent star this presents observers with an important opportunity to resolve, in the time-domain, flux that is either emitted by the planet's atmosphere or flux that is scattered (``reflected'') from the parent star.  Since the discovery of transiting exoplanets \cp{Charb00,Henry00,Borucki10} this opportunity has been seized and we can now detect and interpret this planetary light measured at secondary eclipse \cp[see, e.g.,][]{Winn10,Kreidberg17}.  In the current era of modest signal-to-noise data, approximations in the calculation of model spectra can be made when comparing with observations.  However, as we move towards higher signal-to-noise observations, as will be obtained with the \emph{James Webb Space Telescope} (\emph{JWST}), it is important to critically examine how model calculations are made.  This ensures that robust conclusions can be drawn about exoplanetary atmospheres when model spectra are compared to data.

In this brief Letter we highlight a physical effect that impacts the calculation of model emission and reflection spectra for planets.  As we will show, this effect is largest when the atmospheric scale height is a non-negligible fraction of the planet's radius, meaning that the apparent photospheric radius of the planet can change significantly with wavelength.  This radius effect will impact a wide range of planets from hot Jupiters down to sub-Neptunes, and even some high mean molecular weight atmospheres.  The effect is related to the observation of a transiting exoplanet's \emph{transmission} spectrum, where, when the planet passes in front of its star, it's apparent radius is seen to be a function of wavelength, depending on the wavelength-dependent opacity sources in the planet's atmosphere, as was predicted by several authors \cp{SS00,Brown01,Hubbard01}.

The magnitude of this well-known physical transit effect can be quantified by estimating the number of scale heights ($H$) probed in a transmission spectrum, where $H=k T / \mu m_H g$, where $k$ is Boltzmann's constant, $T$ is the atmospheric temperature at the pressures of interest, $\mu$ is the dimensionless mean molecular weight, $m_H$ is the mass of the hydrogen atom, and $g$ is the planet's surface gravity.  $T$, $\mu$, and $g$ will vary with height in the general case.  While it is now quite well-appreciated by the community that the transit radius of an exoplanet with an atmosphere is indeed a function of wavelength, below we describe the implications of the fact that the same is true for the emitting surface of the planet, which impacts its emission and reflection spectrum.  

A model planet-to-star flux ratio $F$ is written as:
\begin{equation}
F= {\left( \frac{R_{\mathrm p}( \lambda )} {R_{\mathrm s}} \right)}^2   \frac{F_{\mathrm p}(\lambda) d \lambda}{F_{\mathrm s}(\lambda) d \lambda}
\label{ratio}
\end{equation}
where planetary and stellar surface fluxes are denoted by $F_{\mathrm p}$ and $F_{\mathrm s}$, respectively, the radius of the planet is ${R_{\mathrm p}( \lambda )}$, here explicitly shown as a wavelength-dependent quantity, and the radius of the star is $R_{\mathrm s}$.  A common simplifying assumption is that $R_{\mathrm p}$ is assumed constant, although some authors in the literature explicitly show the wavelength dependence for ${R_{\mathrm p}}$ \cp{Drummond18,Gandhi18} \footnote{An unscientific survey of a range of atmosphere modelers (see acknowledgements) found that a wavelength-dependent radius was largely not included, but some workers who did include it did not mention it in their written work, making it difficult to assess the state of the field from the literature alone.}  Given low signal-to-noise data, this is entirely reasonable, even if it is widely appreciated that the photospheric pressure of a planetary atmosphere varies with wavelength.  The question of course is how much $R_p$ varies with wavelength and could changes in the photospheric radius with wavelength be an important physical effect?  It has previously been demonstrated for a wide variety of planetary atmospheres that the photospheric radius varies across several scale heights in pressure.  If these several scale heights reach a significant fraction of the planetary radius, this will impact model emission spectra, following Equation \ref{ratio}.  Below we will examine over how many scale heights the emitting surface of the atmosphere can vary and quantify its impact on secondary eclipse spectra.

\section{Model Calculations}
\subsection{Photospheric Pressure and Radius}
The concept of the photosphere is reasonably well-defined in the stellar atmospheres context as the atmospheric level one sees down to when observing at wavelengths that probe continuum opacity sources.  The notion of any continuum in molecule-dominated atmospheres is much more fraught, and it is better to discuss the photosphere as a function of wavelength.  As a straightforward example we have calculated a model atmosphere and spectrum for WASP-17b, a low gravity hot Jupiter (log $g=2.75$). Our methods have been extensively described previously \cp[e.g.,][]{Fortney05,Fortney08a,Fortney13,Marley15}.

\begin{figure}[hbp!]
\includegraphics[clip,width=1.0\columnwidth]{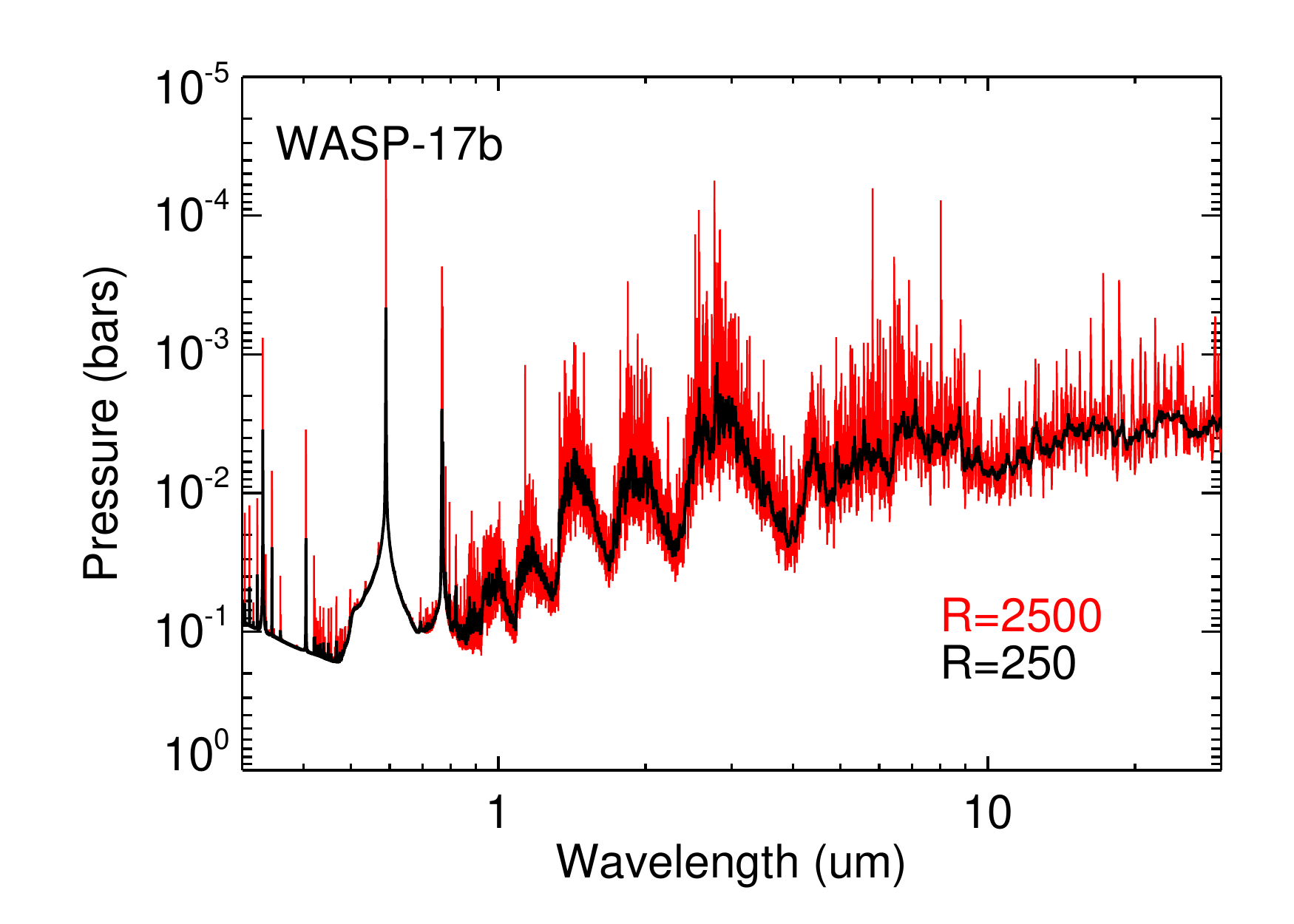}
\caption{For hot Jupiter WASP-17b, the pressure level at which the optical depth $\tau$ reaches $2/3$, which is the wavelength-dependent photospheric pressure.  A calculation at higher resolution ($R=2500$ at 2 $\mu$m) is shown in red, and smoothed to lower resolution ($R=250$ at 2 $\mu$m) is shown in black. The photosphere varies across $\sim$~7 atmospheric scale heights at higher resolution and across $\sim$4.5 scale heights at lower resolution. \label{pressure}}
\end{figure}

Figure \ref{pressure} shows the pressure level in the atmosphere where the optical depth $\tau=2/3$, as a function of wavelength.  This model is shown both at medium resolution ($R=2500$, in red) where the photosphere varies across $\sim$~7 atmospheric scale heights (0.1 bar to 0.1 mbar) and smoothed to lower resolution ($R=250$), across $\sim$4 scale heights (0.1 bar to 2 mbar).  For WASP-17b, where $H=1900$ km, a factor of $7 \times H/R_{\mathrm p}$ is 0.096.  This means that the emitting disk of the planet, which depends on $R_{\mathrm p}^2$, changes by $20.1\%$ across these wavelengths at $R=2500$ (and $12.7\%$ at $R=250$).  While the example of WASP-17b is a hot, low gravity planet, later we will examine the planetary population as a whole. 

\subsection{Effect on Calculated Flux Ratios}
We wish to make clear that this is a wavelength dependent effect and not a simple offset.  The key outcome is that the effect either mutes or enhances features seen in the spectrum.  For example, consider a simple atmosphere where temperature increases with depth.  At wavelengths where one sees deeper into the atmosphere, to a higher brightness temperature, due to low opacity, the planet will appear brighter.  At wavelengths where one cannot see as deeply, due to a strong absorption line or band, the planet will appear dimmer.  However, the \emph{weighting} of the bright wavelengths will be less than the weighting for the dim wavelengths, because the planet's apparent disk is \emph{physically smaller} at the bright wavelengths and is \emph{physically larger} at the dim wavelengths.  This overall mutes the features in the spectrum and one would infer a more isothermal temperature structure than the planet's true structure.

\begin{figure}[htp]
\includegraphics[clip,width=1.0\columnwidth]{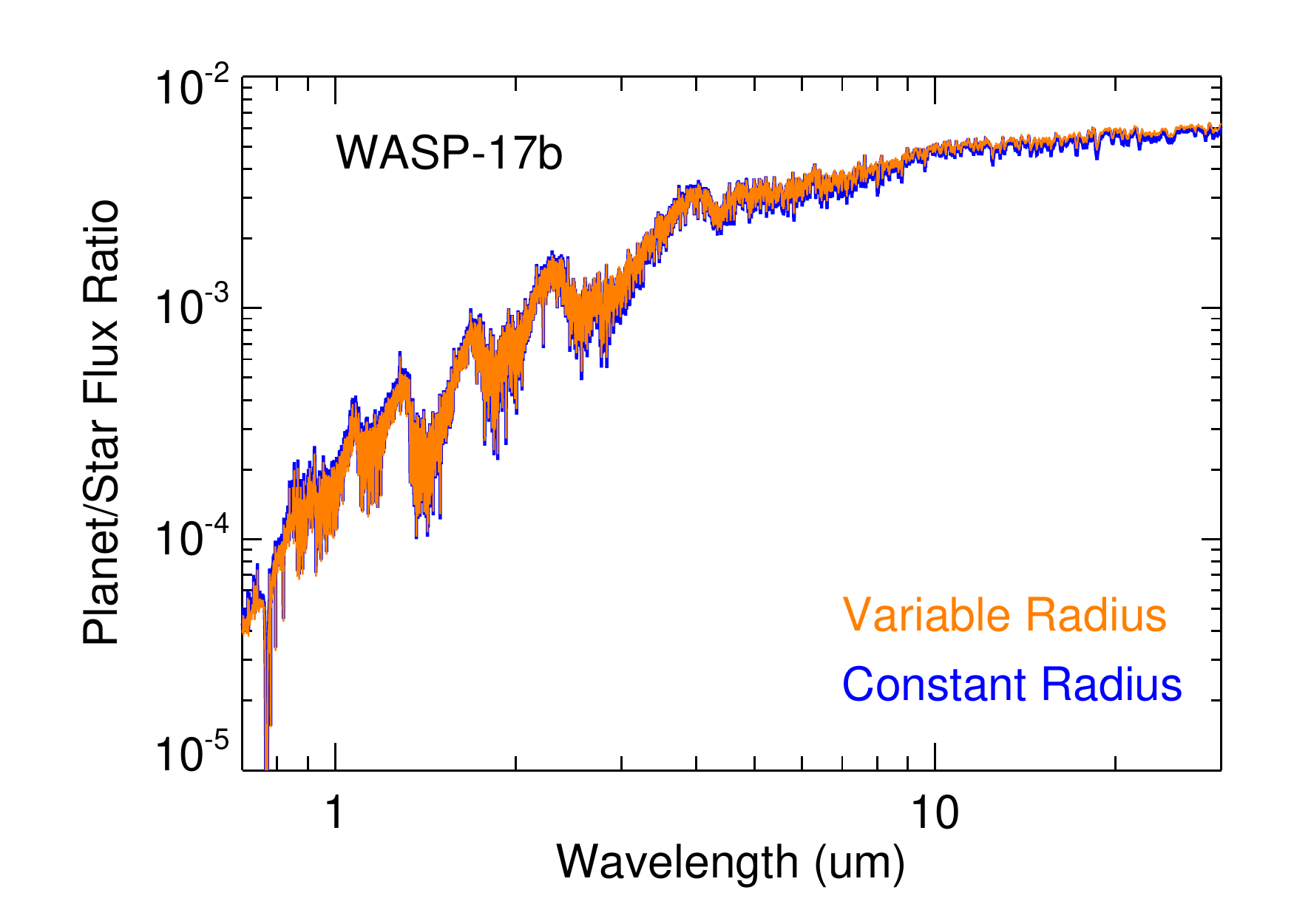}
%\caption{. \label{spec}}
\includegraphics[clip,width=1.0\columnwidth]{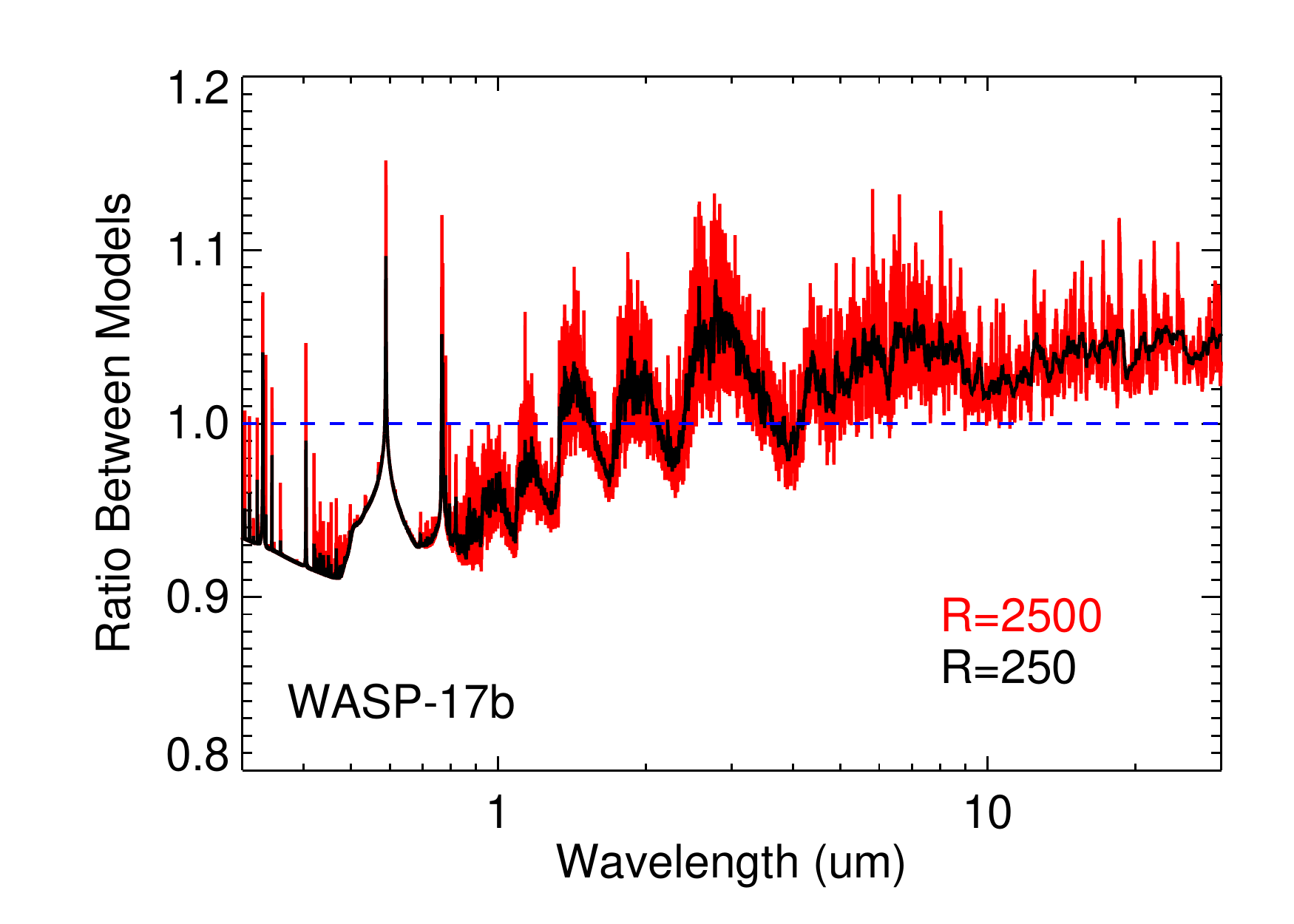}
\caption{Top: Two WASP-17 model planet-to-star flux ratios.  The same model spectra are used for the two flux ratios, but in orange a wavelength-dependent planet radius is used, while in blue is the standard assumption of a constant planet radius.  The spectrum in orange has less dynamic range due to differential weighting caused by the wavelength-dependent planetary radius. Bottom: A ratio plot of the same two models shown at low resolution (black, $R=250$ at 2 $\mu$m) and high resolution (red, $R=2500$ at 2 $\mu$m). Across the infrared, this ratio varies from $\sim$0.95 to 1.05, at 10\% effect at $R=250$. \label{ratioplot}}
\end{figure}

Figure \ref{ratioplot}a shows a calculation of WASP-17 planet-to-star flux ratios.  In blue is a standard calculation where the planet's radius is given from a white-light transit depth.  In orange is a model calculated where the planet's radius changes with wavelength, depending on the depth observed at each wavelength.  The overall effect is to the mute the spectral slope. 
On this log scale, the effect can appear rather subtle, but it becomes clearer when viewed as a ratio between the two calculations, which is shown in Figure \ref{ratioplot}b.  At lower resolution (black), across the infrared, differences reach over 10\% peak-to-trough ($\sim$0.94 to 1.06 in the ratio), and at higher spectral resolution the effect reaches just over 20\% from the near to mid infrared.  This nicely verifies our earlier suggestion that the number of scale heights probed across the photosphere is the dominant factor, which we suggested would be a 20\% effect.  For observational context, $R=2500$ is expected for the \emph{JWST} NIRSPEC instrument from $1-5 \mu$m, while $R=250$ is in the range expected for MIRI LRS ($R=100$) and NIRISS ($R=150-700$) instruments \cp[e.g.,][]{Beichman14}.

At high spectral resolution ($R=30,000-100,000$) from NIRSPEC on Keck or CRIRES on VLT, the effect would be even larger.  However, given that cross-correlation techniques are used \cp[e.g.,][]{Birkby17} at these high resolutions, which focus on the strongest lines \cp{deKok14}, the dynamical range in radius between all the strongest lines would be the proper metric, and would be considerably smaller than the peak to trough.  That being said, including this effect we expect may yield modestly stronger cross-correlations, all things being equal.  In the current space-based observational context (dominated by \emph{Hubble} at very low resolution) this effect would be quite marginal, which is likely why it has often not been implemented.

Of course other atmospheric temperature structures will exist.  When considering an atmosphere with a thermal inversion, the photospheric radius effect is reversed.  More weight is given (a larger emitting area for the planet) for wavelengths that are bright and less weight is given to wavelengths that are dim.  Therefore, features in the calculated spectrum are enhanced compared to the case where a constant planet radius is assumed.  A truly isothermal atmosphere would show a blackbody spectrum no matter how many scale heights over which the photospheric radius changed.

For the case of a reflection spectrum the effect on the calculated spectrum is analogous to that of the non-inverted atmosphere.  Wavelengths with high opacity and less scattered light have a greater weighting due to a physically larger planet.  However, the interpretation would no longer be in terms of an incorrectly inferred temperature structure.  Since the shape of the model spectrum would be slightly incorrect, it would likely be assumed that there were inaccuracies in the underlying opacities used in the model atmosphere, or perhaps even that a thin cloud layer was present, that would subtly mute absorption features.

\section{Range of Applicability}
As suggested above, the magnitude of this photospheric effect is predominantly controlled by $H/R_{\mathrm p}$.  This suggests that the effect will be important for a wide range of planets, if $H$ can be large, due to low surface gravity, high temperature, low mean molecular weight, or a combination of the three.  While this certainly includes hot Jupiters, it will also affect sub-Neptune mass planets as well, where $R_{\mathrm p}$ may be $\sim$3 \re, rather than the 15 \re\ typical of hot Jupiters.  A striking example would be the ``super-puffs'' \cp{Lee16}, planets with radii of that of Neptune or larger, but with masses of only a few Earth masses, such as is found in the Kepler-51 system \cp{Masuda14}.

\begin{figure}[htp]
\includegraphics[clip,width=1.0\columnwidth]{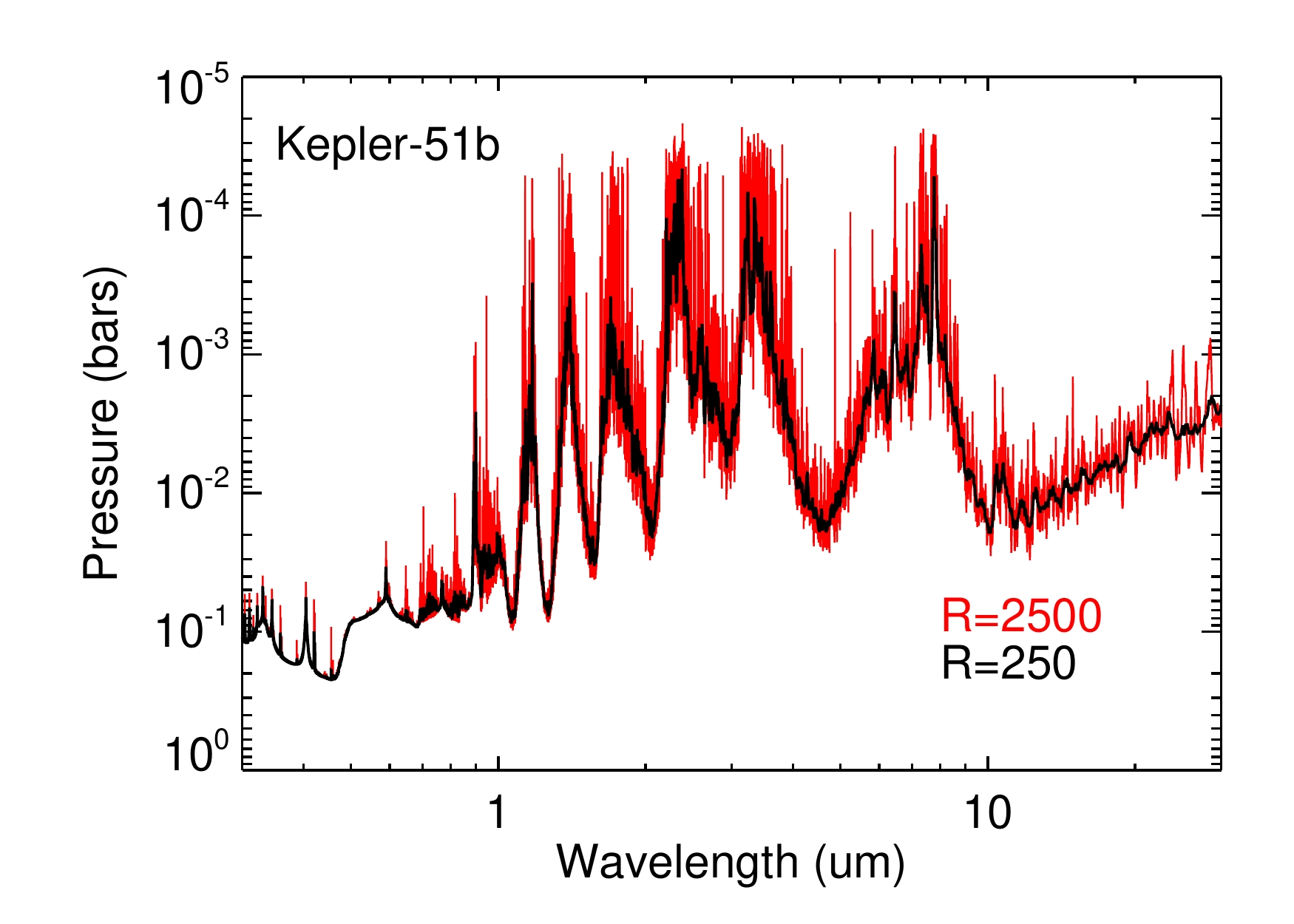}
\includegraphics[clip,width=1.0\columnwidth]{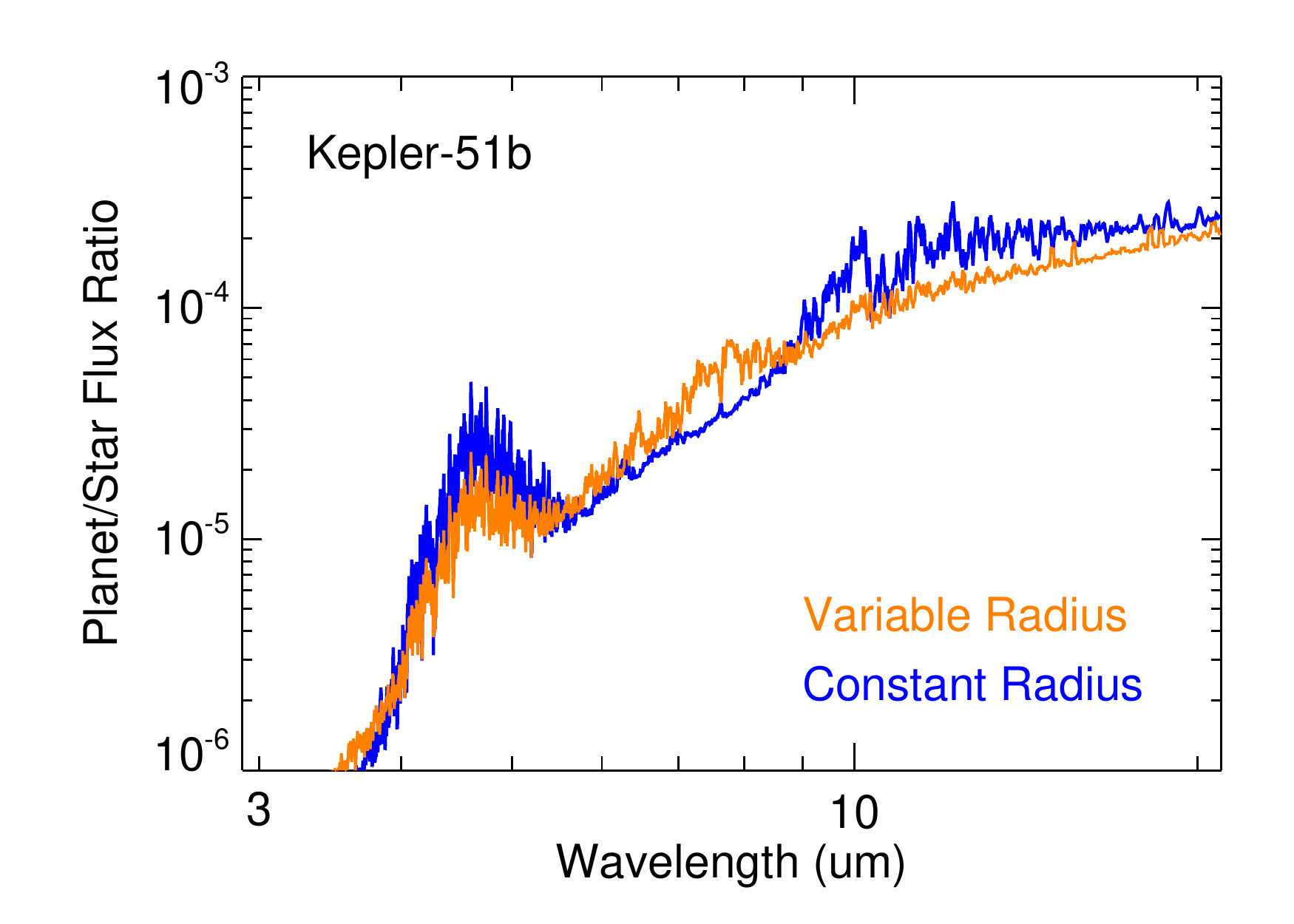}
\includegraphics[clip,width=1.0\columnwidth]{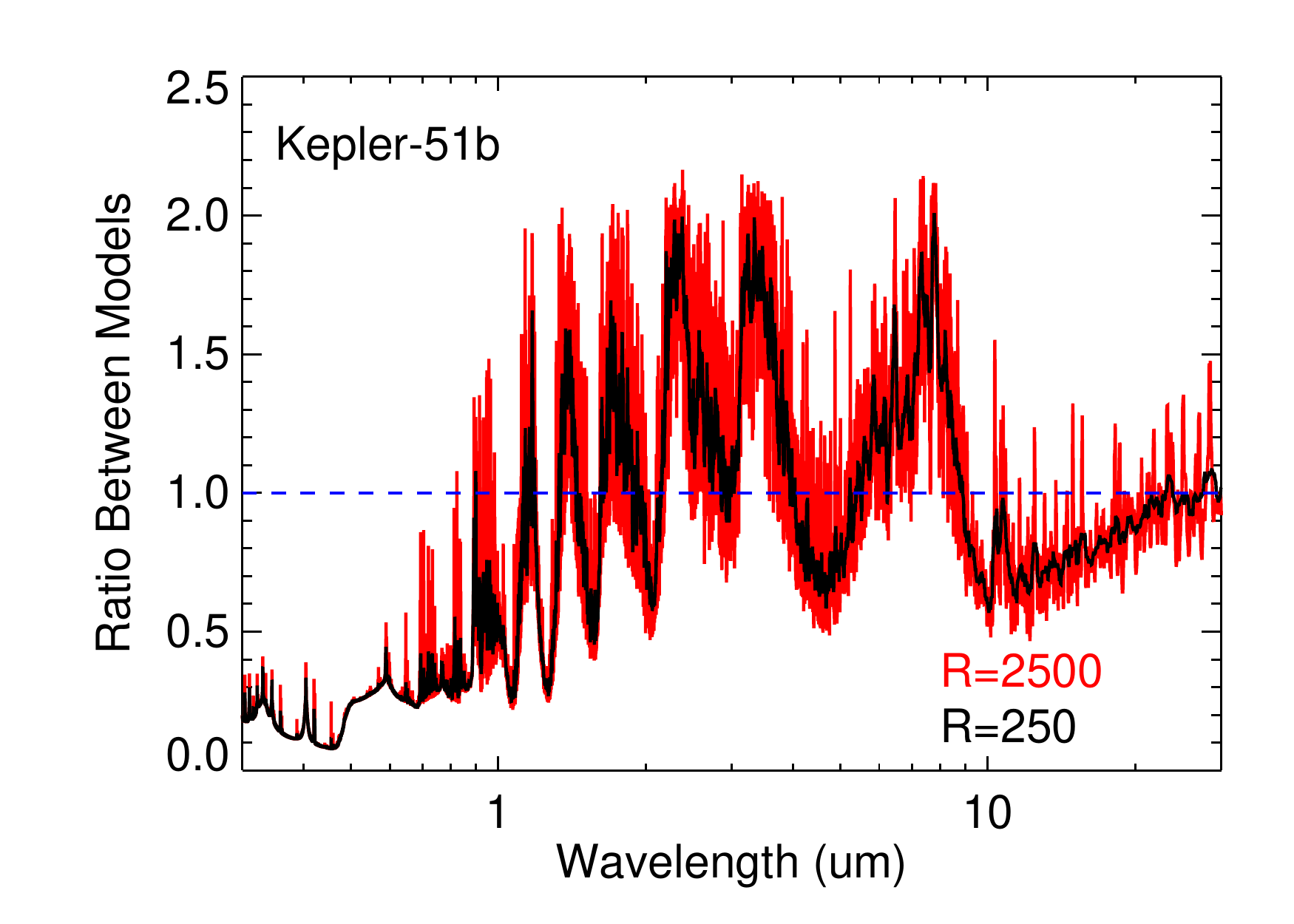}
\caption{Top: The photospheric pressure for a Kepler-51b model, as viewed at two values of spectral resolution.  Middle:  The same model spectra are used for the two flux ratios, but in orange a wavelength-dependent planet radius is used, while in blue is the standard assumption of a constant planet radius.  The spectrum in orange has less dynamic range due to differential weighting caused by the wavelength-dependent planetary radius. Bottom: A ratio plot of the same two models shown at low resolution (black, $R=250$ at 2 $\mu$m) and high resolution (red, $R=2500$ at 2 $\mu$m).  Across the infrared, this ratio varies from 0.3 to 2.0, around 270\%. \label{51plots}}
\end{figure}

Figure \ref{51plots} shows a series of atmosphere calculations for Kepler-51b, at 2.1 \me\ and 7.1\re, an extremely low gravity planet.  These calculations assume a $10\times$ solar atmospheric metallicity.  The top plot shows the range of photospheric pressures.  Compared to WASP-17b, the lower gravity and higher metallicity pushes the photosphere to lower pressure in the strongest bands, but the weakening of the water features outside of the strong bands, due to the lower temperature \cp[e.g.][their Figure 2]{Tinetti12}, leads to a very large dynamic range in pressure, across a factor of 1000 at $R=250$, in black, or 7 scale heights.  Figure \ref{51plots}b shows the resulting emission spectra, in analogy to Figure \ref{ratioplot}a, showing that the model with a wavelength-dependent radius (orange) is significantly more muted in it's features.  Finally, the bottom plot shows the ratio between these two models at two values of spectral resolution.  In analogy for WASP-17b, here $H=4500$ km, log $g=1.6$, $\mu=2.5$, \teq$=550 K$, so that $7 \times H/R_{\mathrm p}$ is 0.69, meaning that the emitting disk of the planet changes by $285\%$ across these wavelengths, even at only $R=250$ (in black).  The variation in the ratio from $\sim$~0.3 to $\sim$~2.0 in Figure \ref{51plots}c, again backs up this straightforward estimate.

\begin{figure}[htp]
\includegraphics[clip,width=1.0\columnwidth]{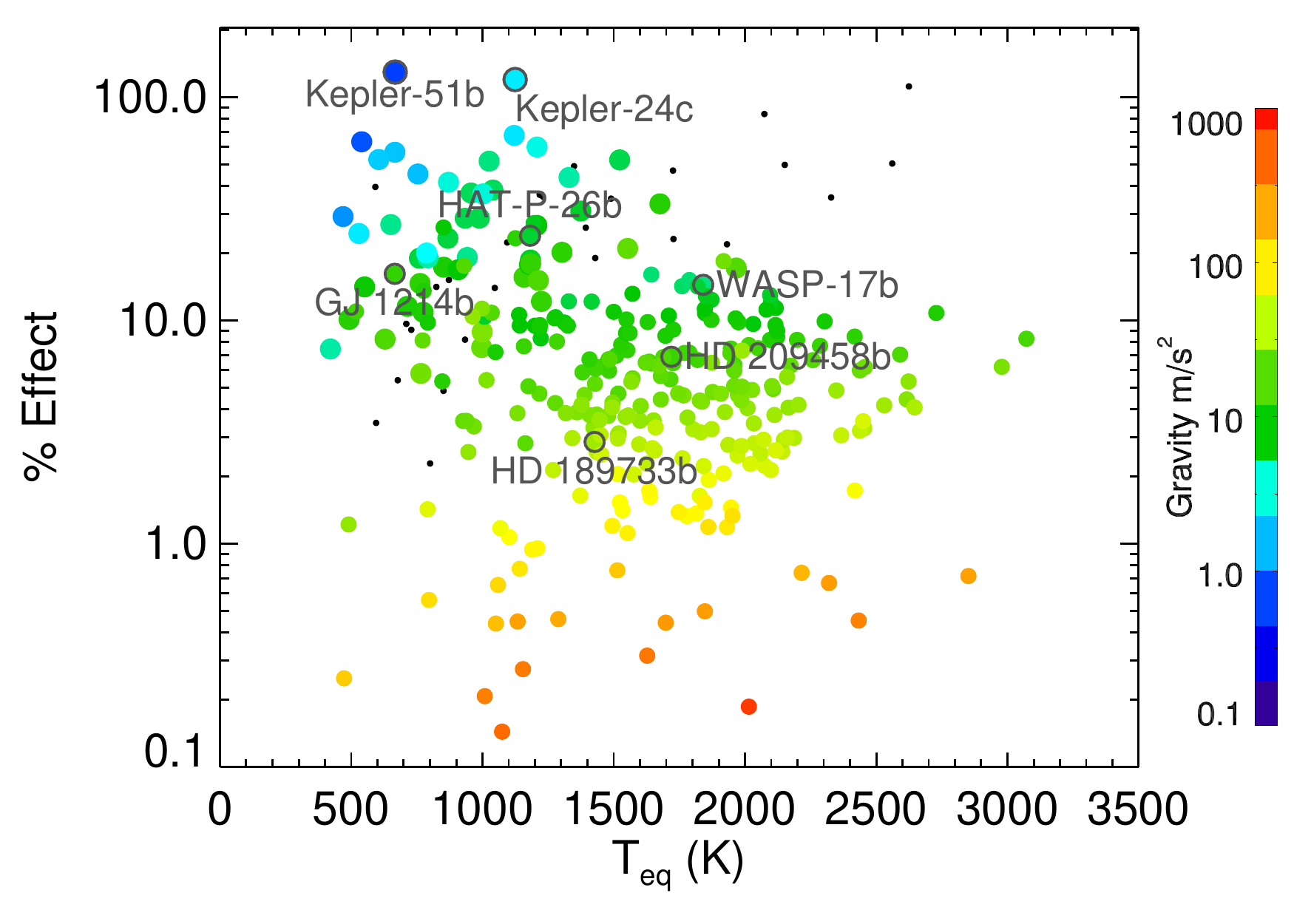}
\caption{An estimate of the size of this photospheric radius effect on planet-to-star flux ratios, as controlled by the change of the ratio of planet-to-star area, for planets larger than 2 \re.  The x-axis is planet \teq, while the color-coding is for planet surface gravity.  The atmospheric dynamic range is assumed to be $4H$, typical of $R\sim~250$ for hot planets, but is an underestimate for the cooler planets shown.  H/He dominated atmospheres are assumed for the calculation, and planets with radii larger than 2.7 \re\ are shown as colored dots, where this assumption is more realistic.  Smaller colored dots are from a compilation from standard exoplanet archives, while larger colored dots are for planets with transit timing variation masses and include ``super puffs" from a catalog of Daniel Jontof-Hutter (personal communication). The effect can reach beyond 10\% for a wide range of planetary temperatures and can reach 50-100\% (or more) for the cooler, but very low gravity puffs. Several planets of interest are labeled in dark gray. \label{HR}}
\end{figure}

Based on our more detailed modeling of WASP-17b and Kepler-51b, we can examine the rest of the exoplanet population.  In Figure \ref{HR} we plot the magnitude of the change in size of the planetary disk assuming a change in $R_{\mathrm p}$ across $4H$, as a function of planetary \teq~ for all transiting planets above 2 \re~with well-determined masses.  This calculation assumes atmospheric $\mu=2.3$ and $T=T_{\mathrm eq}$, with zero Bond albedo and redistribution of stellar flux over the planet's day-side.  Planets larger than 2.7 \re, which will have at least some H/He in their atmospheres, are shown as larger colored dots, with the colors reflecting their surface gravity.  

Planetary atmospheres exist over a wide range of conditions, including those that are only marginally stable at the ``cosmic shoreline'' of escape \cp{Zahnle17} at any value of \teq.  This means that planets need not be hot to have $H$ become a non-negligible fraction of the planetary radius.  For instance, the super-puff planets, shown with larger-size dots in Figure \ref{HR}, are mostly found in orbits cooler than the hot Jupiters.  This photospheric effect need-not only affect models of hydrogen-dominated atmospheres.  For example, even for Saturn's $\sim$~100 K moon Titan, this effect would alter emission spectra over 5 scale heights of photospheric pressure by 8\%, showing that it is not an effect limited to the hottest planets, but to any planet where $H/R_{\mathrm p}$ is not negligible.

\section{Discussion \& Conclusions}
There are several subtleties that make the straightforward application of this effect a bit challenging to implement in calculations.  Importantly, the photospheric radius is not an easily measured quantity.  It is unclear if a measured radius-dependent transit radius (as would be measured from transmission spectrum at primary transit), and the number of scale heights probed in transmission, would adequately determine the number of scale heights probed in emission.  Perhaps most significantly, this difficulty is encountered because the effects of cloud opacity are much more significant at the long paths appropriate for transit spectra \cp{Fortney05c} so that these clouds may not be optically thin for emission spectra, or clouds on the limb may not even be present over the day side, a generic outcome of some 3D simulations \cp{Parmentier16}.  Therefore it seems likely that more scale heights will be probed in emission than transmission.  In fact, using the \emph{measured} transit radius at all wavelengths in Equation 1 will bias one's results towards lower planetary fluxes and temperatures, as the planet's day-side emitting radii will be smaller than the transit radii, at a given wavelength \cp[e.g.][]{Burrows07}.

Our recommendation then is at least for self-consistency:  For the model atmosphere that is generated for the calculation of a spectrum at secondary eclipse, either for 1D radiative-convective equilibrium models, 1D forward models in retrieval, or in the generation of spectra from 3D temperature structures, the corresponding wavelength-dependent photospheric radii from the model should be used in the calculation of the planet-to-star flux ratios \cp[e.g.,][]{Drummond18,Gandhi18}.

In this Letter we have explored the physical effect that the wavelength-dependent photospheric radius of a planetary atmosphere can quantitatively impact the calculation of model planet-to-star flux ratios.  It works to mute features in model spectra from atmospheres that decrease in temperature with height, but to enhance features where there is a thermal inversion.  This effect scales with the ratio of the atmosphere's scale height to the planetary radius, $H/R_{\mathrm p}$, and can be important at any planetary \teq.  For current ``typical" transiting planets this leads to a wavelength dependent correction to model spectra that is around 5\%, but can rise to 15-25\% for low gravity hot Jupiters, and well beyond 50\% for the lowest gravity planets, at $R~\sim250$.  The effect becomes larger at higher spectral resolution, when a higher dynamic range of pressures are probed.

As the exoplanet atmospheres field continues to advance, other physical effects for both planets and stars should be investigated in detail regarding their effects on the calculation of spectra and planet-to-star flux ratios of exoplanetary atmospheres.  For instance, the photospheric radius effect described here could be further investigated with a height-varying gravity falloff, but additional choices would have to be made regarding at what pressure to locate the reference gravity.  A simple choice could be 1 bar, as is typically done for the solar system's gas giants, but this again require an uncertain model fit to a transmission spectrum, as originally pointed out in \ct{Hubbard01}.  We note that the effect shown here is negligible for dwarf parent stars themselves, as both the surface gravities and radii are $\sim$~10$\times$ larger for dwarf stars than for their transiting planets.  Other complexities due to stellar surface features are certainly important, as has recently been investigated for spotty parent stars during spectral observations during the transit \cp[e.g.,][]{Rackham18a,Zhang18,Rackham18b}.  As our observations improve, additional thought should be put into these and similar issues.

\acknowledgements
The authors thank Nathan Mayne, Mike Line, Paul Molliere, Natasha Batalha, Matej Malik, Travis Barman, Bjorne Benneke, Adam Burrows, Kevin Heng, and David Spiegel for fostering interesting discussions.  We thank Eliza Kempton for providing important comments on an earlier draft, Eric Lopez for assistance with figures, and Adam Smith for a close reading.  We thank the referee for a constructive review that improved the presentation of the Letter.  We acknowledge Didier Saumon and Mark S. Marley for contributions to the opacity database used in this paper.

\end{document}